# Photo absorption enhancement in strained silicon nanowires: An atomistic study

Daryoush Shiri[1*a], M. Golam Rabbani[2b], Jianqing Qi[2], Andrei Buin[3c], M. P. Anantram[2]

[1]Institute for Quantum Computing (IQC), Department of Physics and Astronomy, University of Waterloo, Waterloo, N2L 3G1, Ontario, Canada
[2]Electrical Engineering Department, University of Washington, Seattle, WA 98195-2500, USA
[3]Department of Electrical Engineering, University of Toronto, Toronto, M5S 2J7, Ontario, Canada

[*]Corresponding Author: daryoush.shiri@chalmers.se
[a]Daryoush Shiri in now with the Department of Physics, Chalmers University of Technology, SE 412 96 Göteborg, Sweden. [b]M. Golam Rabbani is now with Intel Inc., Portland, Oregon, USA. [c]Andrei Buin is now with D&D Integrative Care Inc., Toronto, Ontario, Canada.

**Abstract:** The absorption spectra of silicon nanowires (SiNW) are calculated using semi-empirical $sp^3d^5s^*$ tight binding (TB) and Density Functional Theory (DFT) methods. The role of diameter, wave function symmetry, strain and crystallographic direction in determining the absorption are discussed. We find that compressive strain can change the band edge absorption by more than one order of magnitude due to change in wave function symmetry. In addition, photon polarization with respect to the nanowire axis significantly alters the band edge absorption. Overall, the band edge absorption of [110] and [100] silicon nanowires can differ by as much as three orders of magnitude. We find that compared to bulk Silicon, a strained Silicon nanowire array can absorb infrared photons (1.1 eV) approximately one hundred times better. Finally, we compare a fully numerical and a computationally efficient semi-analytical method, and find that they both yield satisfactory values of the band edge absorption.

## I. INTRODUCTION

Silicon nanowires (SiNW) have been actively studied both experimentally and theoretically. Compatibility with traditional silicon microelectronic fabrication methods, and demonstrations of quantum confinement, tunable bandgap, and sensitivity of electronic structure to surface adsorbents and mechanical strain, offer significant long-term prospects in applications. During the last decade, we have witnessed applications in transistors[1,2], logic circuits[3], memory[4], spin-based quantum computing[5], chemical[6] and biological[7] sensors, piezo-resistive sensors[8], nano mechanical resonators[9] and thermoelectric converters[10,11]. The utilization of SiNWs in optoelectronic devices such as solar cells[12,13], photo-transistors[14,15], and avalanche photodiodes[16,17] are also promising. Integration of silicon with photonics is important due to low cost and mature fabrication technology of silicon chips. Over the years, researchers have attempted to observe light emission with the signature of quantum confinement from SiNWs. The availability of new fabrication methods[18] has made it possible to reduce non-radiative and surface recombination rates. This is important for light emission arising from transition between conduction and valence band states (as opposed to defect states). At a more fundamental level, light emission/absorption at bandgap energies in SiNWs mandates having a direct bandgap and a symmetry allowed dipole transition between states at the conduction (c) and valence (v) band edges,

$$\langle \Psi_c(r)|r|\Psi_v(r)\rangle = \int \Psi_c^*(r) r \Psi_v(r) d^3r \qquad (1)$$

To have a non-zero (symmetry allowed) value for the dipole moment, the integrand in equation (1) should be of even symmetry. Here, $\Psi_c$ and $\Psi_v$ are the conduction and valence band wave functions, and **r** is the position operator. In bulk silicon and large diameter SiNWs, the conduction band minimum and valence band maximum do not have the same wave vector value within the Brillouin Zone (BZ) [Figure 1(a)]. Since the photon absorption/emission is a momentum conserving process, indirect transitions are weak second order processes mediated by phonons. However, narrow diameter SiNWs can have a direct bandgap due to folding of off center energy states of bulk silicon into the BZ center. Figure 1(b) shows a direct band gap for a [100] SiNW with a diameter of 2.2 nm. However, most experiments have involved large diameter SiNWs within a hybrid light emitting device, where photon emission occurs efficiently via a direct bandgap nanowire made from III-V materials. In these demonstrations, the SiNW is used only for biasing and carrier injection. For example, nano-lasers and Light Emitting Diodes (LED) were reported based on the combination of top-down grown SiNWs and, bottom-up III-V[19] and Cd-VI nanowires[20,] as well as a core/shell p-Si/n-CdS nanowire [17]. Further, studies utilizing SiNW arrays in solar cells, wave guides[21], color filters[22] and photonic crystals exploit the electromagnetic wave scattering properties of these arrays rather than the quantum confined properties of SiNW because their diameters are between few tens of nanometers and a 1µm. The role of SiNW arrays in these experiments are to help with (a) Spectrum Widening and (b) Collimation and Lensing. *Spectrum Widening* refers to increase in optical absorption of a SiNW array over a wider wavelength range than thin film silicon of equivalent thickness[23-26]. *Collimation and lensing* refers to the ability of a periodic nanowire array to collimate IR (λ = 1500 nm) wavelengths to increase absorption[27,28]. Finally, there are examples of prototype devices using a single large diameter SiNW as photodiodes, solar cells and photo transistors. A detailed review of these experiments and the performance of devices can be found elsewhere[29].

The role of strain in engineering electronic and optoelectronic properties has proved important in bulk materials (e.g. Si, Ge). The residual stress due to lattice mismatch between silicon and germanium can enhance the carrier mobility in Metal-Oxide-Semiconductor (MOS) transistors. Additionally, it has been demonstrated that growing a germanium layer on a silicon substrate can decrease the splitting between the indirect and direct bandgap in germanium[30,31]. The reduction in splitting in turn increases the light emission efficiency of optically[32] and electrically pumped[33] Ge-on-Si. On the computational side, using the TB method, it was shown that a +2% strain value changes the bandgap of bulk Ge from indirect to direct[34]. This value of strain is similar to the value of +1.8% predicted by the effective mass theory[31] as well as the experimental value of +2% reported in Ge-on-Si light emitting diodes[32,33]. The value of strain in these thin films is fixed because it is controlled by alloy composition[35,36]. Similarly, coaxial SiNWs can be strained by a lattice mismatch induced by the shell material; examples include germanium[37], $SiO_2$[38-40,18], silica[42], metallization[2] and silicon nitride[43,44]. Strain was also found to modulate the spontaneous emission rate in narrow diameter SiNWs by two orders of magnitude[45] as a result of symmetry change in wave functions and a direct to indirect bandgap transition[46-53]. It was recently observed that the electron-hole recombination rate in silicon nanowires can be enhanced by 6 times due to strain (1%-5% tensile) induced in the bent sections of the nanowires[54].

Motivated by the effect of strain in modulating the spontaneous emission rate, the main goal of this article is the computational study of photo absorption in SiNWs and its dependence on strain, diameter, and crystallographic direction. Due to the equivalence of the stimulated emission and absorption rates ($R_{stim}=R_{abs}$), this study facilitates the calculation of gain spectrum in SiNW-based lasers as well. We start from the quantum mechanical formulation of the dielectric function of semiconductors adapted for a nanowire by including 1D Joint Density of States (JDOS) using the ten orbital ($sp^3d^5s^*$) semi-empirical tight binding method[55]. We note that bandgap underestimation in DFT makes direct comparison of light absorption in SiNWs with bulk silicon less straightforward. On the other hand, TB has already shown success in reproducing the band structure of bulk silicon as well as the experimental strain-induced shift of PL spectra in SiNWs[56,18].

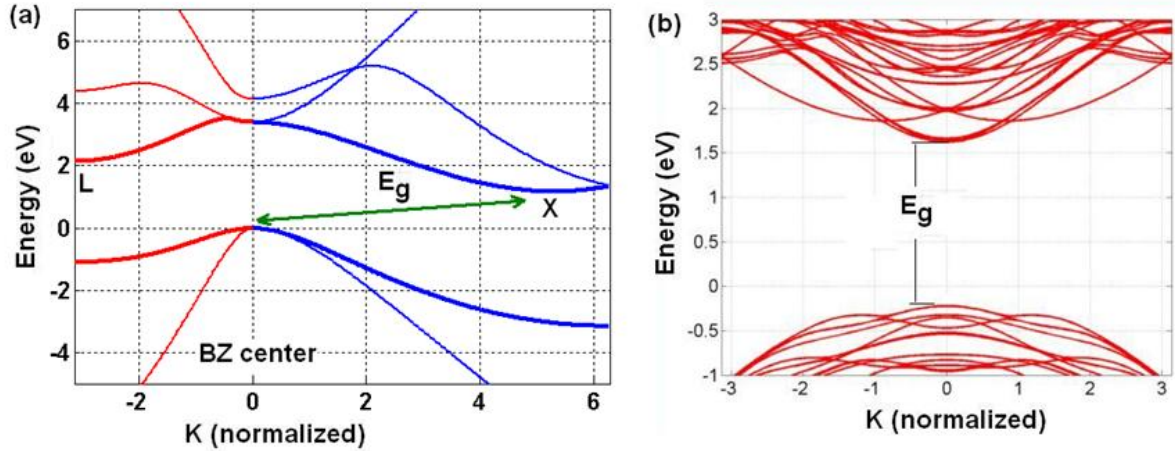

FIG. 1. (a) Band structure of bulk silicon showing an indirect bandgap. (b) Band structure of a [100] SiNW with a diameter of 2.2 nm shows a direct bandgap in contrast to bulk silicon.

The absorption spectrum can be found using both real and imaginary parts of the dielectric function. The real part of dielectric function is calculated from the imaginary part using the Kramers-Krönig relationship. The approximations can be summarized as follows: (a) the tight binding method is used, and is validated by comparing with DFT[57] and experiments in case of bulk Silicon. (b) The optical transitions studied in this works are all of direct type (first order) and indirect (second order phonon mediated) transitions are not studied here. (c) After observing close agreement of the full band (fully numerical) and effective mass (semi analytic) approximations, the latter is chosen. The transition matrix element value is assumed to be close to the value at the Brillouin zone center. (d) The local field effect and many-body effect are not considered in this work. (e) The calculations are carried only to linear response. (f) In evaluating the matrix elements, it is assumed that the orbitals of Silicon and Hydrogen are of Slater type. In carrying out the validation mentioned in (a), we use the DFT method implemented in SIESTA[57] to calculate the absorption spectrum of bulk silicon only due to direct transitions. Calculation of the absorption in indirect bandgap semiconductors has been discussed in the literature[58] and a recent DFT study reports the calculation of absorption spectrum (phonon assisted part) in bulk silicon using the second order perturbation theory[59].

This article is organized as follows. We start with a review of computational methods i.e. energy minimization with and without strain, and band structure calculations using the tight binding method (II.A). This is followed by a comparison of a fully numerical and a computationally efficient semi-analytical computation of absorption spectrum using the TB method (II.B, B.1). The results from these computations are compared to DFT-based and experimental absorption spectra for bulk silicon (II.B.2). Then, the effects of diameter, crystallographic direction and strain value on band edge absorption of SiNWs are discussed along with an explanation of the underlying principles (section III). Section IV concludes the article.

## II. METHODS

**A. Energy Minimization and Strain Application:** The SIESTA package[57] is used to relax the atomic structure of periodic silicon nanowires[60]. The Local Density Approximation (LDA) functional with Perdew-Wang (PW91) exchange correlation potential[61] is used, along with spin polarized Kohn-Sham orbitals of double-ξ type with polarization. The Brillion Zone is sampled by a set of 1×1×40 k points along the nanowire axis (z axis). The minimum center to center distance of SiNWs is at least 6 nm to avoid inter-cell interaction. The values for energy cut-off, split norm and force tolerance are 680 eV (50 Ry), 0.15, and 0.01 eV/Å, respectively. The energy of the unstrained nanowire is minimized using Conjugate Gradient (CG) algorithm during which the variable unit cell option is used. An initial guess for the strained nanowire is generated by uniformly stretching the unstrained nanowire according to $a_{new} = a_{old}(1 \pm \varepsilon)$, where a is the unit cell length and ε is strain value in percent [see Figure 2(a)]. At nonzero values of strain, the unit cell is relaxed using the constant volume (fixed unit cell) option.

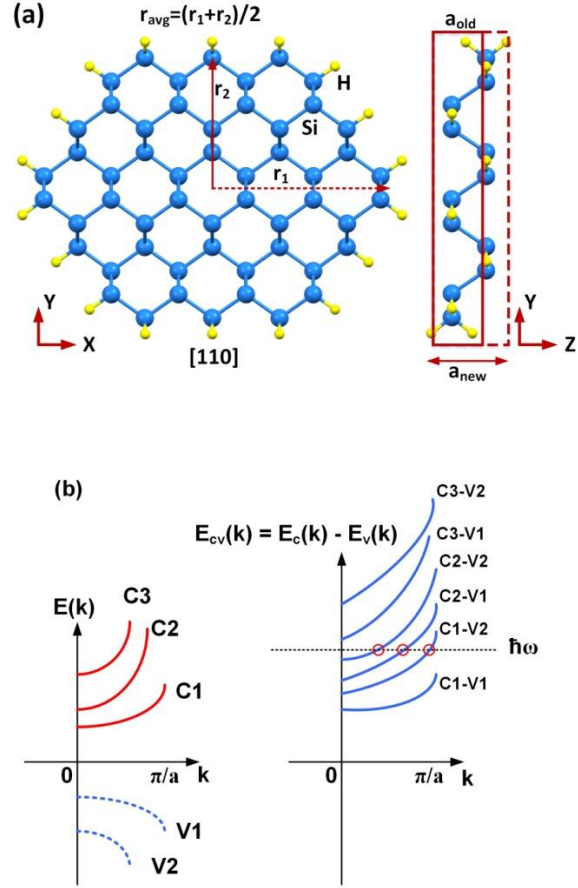

FIG. 2. (a) Illustration of a [110] SiNW unit cell with diameter of 1.7nm. The diameter is defined as the average of large and small diameters. Dark and bright atoms correspond to Si and H, respectively. Application of strain is performed by updating the old unit cell ($a_{old}$) by the given strain percent ($\varepsilon$). (b) Pictorial representation of finding/counting the number of available inter-band optical transitions between valence (v) and conduction (c) bands for a given photon energy $\hbar\omega$.

We calculate the band structure and Eigen states using the ten orbital ($sp^3d^5s^*$) TB method with Jancu's parameters[55]. The bandgap underestimation that is typical of DFT calculations is avoided by using the TB method. The TB method has successfully reproduced the effect of radial strain on the PL spectrum of narrow SiNWs in the experiments[56,18]. The trend of TB bandgap change with diameter in the case of silicon nano-crystals also agrees with DFT based calculations[62]. In the case of [110] SiNWs, the TB method reproduces the bandgap change with diameter observed in Scanning Tunneling Spectroscopy (STS) measurements[63]. Finally, we note that the nanowires in this work are constructed from a bulk silicon crystal in [110] and [100] directions and terminating their surface with hydrogen atoms. The cross section of each nanowire lies in the x-y plane and its axis is parallel to z direction. Figure 2(a) shows the atomic structure of a unit cell in a SiNW oriented in [110] direction with a diameter of 1.7 nm.

**B. Photo absorption**

The calculation of the tensor dielectric function ($\epsilon_{\alpha\beta}$) of the material is required to obtain the absorption spectrum. Macroscopically, the displacement vector (**D**) and applied electric field (**E**) are related by:

$$\boldsymbol{D}_\alpha(\boldsymbol{q},\omega) = \epsilon_{\alpha\beta}(\boldsymbol{q},\omega)\boldsymbol{E}_\beta(\boldsymbol{q},\omega). \quad (2)$$

ω and **q** are frequency and wave vector of the electromagnetic field. For anisotropic solids (non-cubic crystals), the non-diagonal elements of $\varepsilon_{\alpha\beta}$ are nonzero. In addition to the above constitutive equation, the electric field induces a current density ($J$) in the material, which is given by,

$$J_\alpha(\boldsymbol{q},\omega) = -\frac{i\omega}{4\pi}[\epsilon_{\alpha\beta}(\boldsymbol{q},\omega) - \delta_{\alpha\beta}]E_\beta(\boldsymbol{q},\omega). \quad (3)$$

$\delta_{\alpha\beta}$ is the Krönecker delta function. The quantum mechanical treatment to calculate the dielectric function of solids was initially demonstrated in Refs. 64-66. Later, this method was used for calculation of optical properties of bulk copper[67] and silicon nano crystals[68]. The procedure is summarized here. The single particle Hamiltonian of the periodic solid with the time-dependent magnetic (**A(r,t)**) and electric (V(**r**,t)) potentials is,

$$\mathcal{H} = \frac{1}{2m}[\mathbf{p} + \left(\frac{e\mathbf{A}(\mathbf{r},t)}{c}\right)]^2 + V(\boldsymbol{r},t) + U(\boldsymbol{r}). \quad (4)$$

The unperturbed periodic potential in the crystal is U(**r**). Since the laser power used in the photoluminescence (PL) studies of SiNWs are typically in the kW/cm$^2$ range, it is possible to ignore the nonlinear perturbation terms (A$^2$), and retain only the linear terms. The response of electrons in the solid to the perturbation results in a change in charge ($\rho(r,t)$) and current densities. The charge and current densities are in turn related by continuity equation, $\nabla.J + \partial\rho/\partial t = 0$. The change in charge and current densities from equilibrium can be obtained from knowledge of the density matrix ($\wp$), which is obtained by solving Liouville equation,

$$i\hbar\, \partial\wp/\partial t = [\mathcal{H},\wp]. \quad (5)$$

The density matrix can be written as, $\wp = \wp^0 + \wp^1$, where the first and the second term correspond to the unperturbed and perturbed density matrix, respectively. The unperturbed part is given by:

$$\wp^0|\boldsymbol{k}l\rangle = f_0(E_{kl})|\boldsymbol{k}l\rangle, \quad (6)$$

where $f_0$ is the Fermi function. **k** and *l* are quantum numbers that represent the electron wave vector and band index of the unperturbed wave function,

$$|\mathbf{k}l\rangle = \frac{1}{\sqrt{V}}e^{ik.r}u_{kl} \quad (7)$$

with energy Eigen value $E_{kl}$. V is the volume of the crystal and $u_{kl}$ is periodic part of the wave function. From the first order perturbed density matrix ($\wp^1$), the induced charge density can be found using:

$$\rho^{ind}(\mathbf{r},t) = Trace(\wp^1)\rho^0_{op}(\mathbf{r}) \quad (8)$$

$\rho^0_{op}(\mathbf{r})$ is quantum mechanical charge density operator, $\rho^0_{op}(\mathbf{r}) = e\delta(\mathbf{r} - \mathbf{r}_e)$, where $e$ and $r_e$ represent charge of electron and its position in the crystal, respectively. Inserting $\wp$ into equation (5) and multiplying it from right (left) by $|\mathbf{k}l\rangle$ ($\langle\mathbf{k} + \mathbf{q}\, l'|$) makes it possible to find $\wp^1$ and $\rho^{ind}(\mathbf{r},t)$. The band index of the new state is shown by *l'* and wave vector change of the electron is due to absorption/emission of a photon with a wave vector of **q**. Using $\rho^{ind}(\mathbf{r},t)$ and continuity equation, the value of $J^{ind}(\mathbf{r},t)$ is known at this stage. Finally, ε(**q**,ω) can be calculated by using equation (3) from knowledge of $J^{ind}(\mathbf{r},t)$. If it is assumed that the wavelength of the incoming light is larger than the unit cell length ($U_c$) of the solid (λ>>$U_c$), then it can be said that q=2π/λ is zero and the dielectric tensor is given by:

$$\varepsilon(0,\omega) = 1 - \frac{4\pi e^2 N}{mV\omega^2} + \frac{4\pi e^2}{m^2 V\omega^2}\sum_{ll'k}\frac{|\langle lk|p|l'k\rangle|^2(f_0(E_{l'k})-f_0(E_{lk}))}{\hbar\omega+E_{l'k}-E_{lk}+i\eta}. \quad (9)$$

Here, N is the number of unit cells in the crystal and η is the broadening factor. From now on, we replace *l* and *l'* with v and c which represents quantum numbers of a state in valence and conduction band, respectively. In the process of obtaining equation (9) the Local Field Effect (LFE) was ignored i.e. it was assumed that there is no distinction between macroscopic and microscopic potentials and electric fields. Otherwise a microscopic potential, $V^{ind}(\mathbf{r}, t)$, is induced from the macroscopic excitation which is V(**r**, t). Hence before starting the previous procedure, e.g. equation (8), $V^{ind}(\mathbf{r}, t)$ must be first calculated from V(**r**, t) using perturbation theory. Later on $V^{ind}(\mathbf{r}, t)$ enters the main procedure as a new perturbation (as if LFE was not included).

In addition to linear response, the following assumptions were made in our calculations: (a) Nanowire is assumed to be an isotropic media i.e. **D** is always parallel with **E**. This is because for cubic crystals like silicon the dielectric tensor is isotropic. (b) Although equation (9) is general and it includes both intra and inter sub band transitions, only the inter sub band transitions are considered for SiNWs [See Figure 2(b)]. This is because the bandgap of SiNWs is large and at room temperature the conduction band is empty. However, in case of a small bandgap material e.g. bismuth nanowire[69] with $E_g$ = 123 meV, the conduction band has significant number of electrons and transitions between conduction band states ($C_1$ to $C_2$ and $C_1$ to $C_3$ etc) cannot be ignored at room temperature [Figure 2(b)]. (c) As a result of the previous assumption (b), the conduction band is empty and the valence band is full i.e. $f_0(E_{vk}) = 0$ and $f_0(E_{ck}) = 1$ (This is equivalent to assuming T= 0°K.).

Finding the imaginary part of ε(0,ω) (equation (9)) is straightforward. Converting CGS to SI unit (multiplying to $1/4\pi\varepsilon_0$) and using the Lorentzian approximation of Dirac's delta function (equation (10)),

$$\delta(\hbar\omega + E_{ck} - E_{vk}) = \frac{1}{\pi}\frac{\eta}{(\hbar\omega + E_{ck} - E_{vk})^2 + \eta^2} \quad (10)$$

results in equation (11)[70-71],

$$\varepsilon_2(\omega) = imag(\varepsilon(0,\omega)) = \frac{\pi e^2}{\varepsilon_0 m^2 V \omega^2} \sum_{cvk} |\langle u_{kv}|\hat{e}.p|u_{ck}\rangle|^2 \delta(\hbar\omega + E_{ck} - E_{vk}). \quad (11)$$

$\hat{e}$ is a unit vector along the direction of photon polarization and $\varepsilon_0$ is dielectric permittivity of vacuum. Now the momentum matrix element between a conduction band state (c, $k'$) and a valence band state (v, $k$) can be written as $\langle u_{kv}|p|u_{ck'}\rangle$ or $\langle v, k|p|c, k'\rangle$. Using the time evolution of the position operator, $\frac{\partial}{\partial t}\mathbf{r} = \frac{i}{\hbar}[H, \mathbf{r}]$ [58], the momentum matrix element can be written in terms of position operator as follows. The Hamiltonian operator acts on each state according to $H|n, \mathbf{k}\rangle = E_{n,\mathbf{k}}|n, \mathbf{k}\rangle$ and therefore:

$$\langle v, \mathbf{k}|\mathbf{p}|c, \mathbf{k}'\rangle = \left\langle v, \mathbf{k}\left|m_0\frac{\partial}{\partial t}\mathbf{r}\right|c, \mathbf{k}'\right\rangle = \frac{im_0}{\hbar}\langle v, \mathbf{k}|H\mathbf{r} - \mathbf{r}H|c, \mathbf{k}'\rangle = \frac{im_0}{\hbar}(E_{v,\mathbf{k}} - E_{c,\mathbf{k}'})\langle v, \mathbf{k}|\mathbf{r}|c, \mathbf{k}'\rangle. \quad (12)$$

Now, the momentum matrix element in equation (11) can be replaced by

$$\left|\langle u_{v,k}|p|u_{c,k'}\rangle\right|^2 = m^2 \omega_{cv}^2 |\langle u_v|\mathbf{r}|u_c\rangle|^2 \quad (13)$$

where $\omega_{cv} = (E_{v,\mathbf{k}} - E_{c,\mathbf{k}'})/\hbar$.

To find the position matrix element in equation (13), we use Slater-type[34] orbitals, which have been successfully applied to semiconductors[72-75] and molecules[76]. We start with an electronic Eigen-state given at a point along the 1D BZ of SiNW which is expanded using the 10 orbital basis. The number of Eigen values for each wave vector is $N_{orbit} = N_{Si} \times 10 + N_H$, where $N_{Si}$ and $N_H$ are number of silicon and hydrogen atoms within each unit cell, respectively. The position matrix element in equation (13) can be reduced to two terms[34,72,76],

$$\langle v, \mathbf{k}|\mathbf{r}|c, \mathbf{k}'\rangle = \sum_{\alpha\beta} C_{v\alpha}^*(\mathbf{k})C_{c\beta}(\mathbf{k}')\delta_{\mathbf{k},\mathbf{k}'}\langle \alpha(\mathbf{r})|\mathbf{r}|\beta(\mathbf{r})\rangle + \sum_{\alpha} C_{v\alpha}^*(\mathbf{k})C_{c\alpha}(\mathbf{k}')\delta_{\mathbf{k},\mathbf{k}'}\mathbf{R}. \quad (14)$$

The coefficients ($C$) are components of each eigenvector, i.e. $N_{orbit}$ complex numbers in general. The indices of $C$, i.e. α and β, span the 10 orbitals for Si and one orbital for H. The second term in the right hand side of equation (14) is simply the element by element multiplication of valence and conduction eigenvectors which are weighted by coordinates of the atoms $\mathbf{R}$ in a unit cell. The first term on the right hand side of equation (14) contains the overlap integral between different types of orbitals on the same atom, i.e. $\langle \alpha(\mathbf{r})|\mathbf{r}|\beta(\mathbf{r})\rangle$ and it is called intra-atomic contribution. This quantity is an integral similar to equation (1) in which both radial and angular parts of each orbital must be used to find the overlap value. Since many of the overlap integrals vanish due to symmetry, only 15 out of the 100 overlap integrals are non-zero and need to be calculated. In this work we assume Slater type orbitals[77],

$$\varphi(r) = N\, r^{n^*-1} e^{-\frac{Z-s}{n^*}r}. \quad (15)$$

Z is the atomic number, $s$ is a screening constant and $n^*$ is effective quantum number. N is the normalization factor which is calculated by normalizing the whole orbital (both angular and radial parts included). We verified that our orbitals match with the orbitals reported in Ref. 75 for the more restricted sp$^3$s$^*$ TB method. However, here we also include 3d and 4s* orbitals following the procedure in Ref. 78 wherein the authors used d* for III-V semiconductors. Table I lists the values of $n^*$, $Z-s$, radial and angular parts of orbitals calculated for a silicon atom.

| Orbital index | Orbital type | n* | Z-s | Radial and angular part |
|---|---|---|---|---|
| 1 | s | 3 | 4.15 | 0.37032 $r^2 e^{-1.383r}$ |
| 2 | $p_x$ | 3 | 4.15 | 0.37032 $r^2 e^{-1.383r}$ sinθcosφ |
| 3 | $p_y$ | 3 | 4.15 | 0.37032 $r^2 e^{-1.383r}$ sinθsinφ |
| 4 | $p_z$ | 3 | 4.15 | 0.37032 $r^2 e^{-1.383r}$ cosθ |
| 5 | $d_{xy}$ | 3 | 1 | 0.0049 $r^2 e^{-0.3333r}$ sin$^2$θsin2φ |
| 6 | $d_{yz}$ | 3 | 1 | 0.0098 $r^2 e^{-0.3333r}$ sinθcosθsinφ |
| 7 | $d_{zx}$ | 3 | 1 | 0.0098 $r^2 e^{-0.3333r}$ sinθcosθcosφ |
| 8 | $d_{x^2-y^2}$ | 3 | 1 | 0.0049 $r^2 e^{-0.3333r}$ sin$^2$θcos2φ |
| 9 | $d_{3z^2-r^2}$ | 3 | 1 | 0.0028 $r^2 e^{-0.3333r}$ (3cos$^2$θ-1) |
| 10 | s* | 3.7 | 1.45 | 0.49×10$^{-4}$ $r^{2.7} e^{-0.3919r}$ |

Table I Slater type orbitals for silicon atom.

The values of fifteen non-zero overlap integrals for silicon are given in Table II. Integration over the radial part is found by numerical integration. For the angular part, the exact integration of spherical harmonics is found analytically[79]. The angular part of the orbitals or $Y_{lm}$, which are written in Table I, can be found in literature[80]. In our case *l* spans 0, 1 and 2, and m spans 0 for *s* and *s*$^*$ orbitals, {-1,0,+1} for three *p* orbitals and {-2,-1,0,+1,+2} for five *d* orbitals.

| Non-zero overlap integrals | value (Å) |
|---|---|
| $\langle s\|z\|p_z\rangle = \langle s\|x\|p_x\rangle = \langle s\|y\|p_y\rangle$ | 1.4636 |
| $\langle s^*\|z\|p_z\rangle = \langle s^*\|x\|p_x\rangle = \langle s^*\|y\|p_y\rangle$ | 0.3697 |
| $\langle p_z\|z\|d_{3z^2-r^2}\rangle$ | 0.4029 |
| $\langle p_x\|z\|d_{zx}\rangle = \langle p_y\|z\|d_{yz}\rangle$ | 0.3525 |
| $\langle p_x\|x\|d_{x^2-y^2}\rangle = \langle p_y\|x\|d_{xy}\rangle = \langle p_z\|x\|d_{zx}\rangle$ | 0.3525 |
| $\langle p_y\|y\|d_{x^2-y^2}\rangle = \langle p_x\|y\|d_{xy}\rangle = \langle p_z\|y\|d_{yz}\rangle$ | 0.3525 |

Table II Non-zero overlap integrals for Slater type orbitals in silicon.

**B.1 Numerical Method:** The one-dimensional band structure of SiNW allows further simplification of equation (11) for numerical calculation. In case of a nanowire, the wave vector ***k*** is a 1D vector along the axis of the SiNWs (z-axis) i.e. $\mathbf{k} = k\hat{z}$. The summation over **k** is reduced to:

$$\sum_k \ldots = 2 \times \frac{L_{nw}}{2\pi} \int_{1D\,BZ} \ldots dk \quad (16)$$

Where $L_{nw}$ is the length of the nanowire and the extra factor of 2 takes the spin degeneracy into account. Let $E_{cv}(\mathbf{k}) = E_c(\mathbf{k}) - E_v(\mathbf{k})$. Now Dirac's delta function can be expanded using the roots ($K_{zp}$) of the equation $E_{cv}(\mathbf{k_{zp}}) - \hbar\omega = 0$. That is,

$$\delta(E_{cv}(k) - \hbar\omega) = \sum_{K_{zp}} \frac{\delta(k - K_{zp})}{\left|\frac{\partial E_{cv}(k)}{\partial k}\right|}. \quad (17)$$

Using equation (17), the dielectric function [equation (11)] can be written as:

$$\epsilon_2(\omega) = \frac{e^2}{\epsilon_0 m_0^2 A_{nw}\omega^2} \sum_{c,v} \int_k |\langle u_c|\hat{e}.\mathbf{p}|u_v\rangle|^2 \sum_{K_{zp}} \frac{\delta(k - K_{zp})}{\left|\frac{\partial E_{cv}(k)}{\partial k}\right|} dk. \quad (18)$$

By interchanging the integration and the last summation and using the sifting property of Dirac's delta, equation (18) becomes

$$\epsilon_2(\omega) = \frac{2e^2}{\epsilon_0 m_0^2 A_{nw}\omega^2} \sum_{c,v} \sum_{K_{zp}\geq 0} \frac{|\langle u_c|\hat{e}.\mathbf{p}|u_v\rangle|^2}{\left|\frac{\partial E_{cv}(k)}{\partial k}\right|_{@K_{zp}\geq 0}}, \quad (19)$$

where $A_{nw}$ is the cross-sectional area of the nanowire. From the calculation of the band structure and Eigen states, all possible combinations of valence to conduction transitions (i.e. $E_{cv}$ and corresponding matrix elements $\langle u_c|\hat{e}.\mathbf{p}|u_v\rangle$) are found for each given *k* value along the BZ ($k_z$-axis) [See Figure 2(b)]. The method of implementing equation (19) is depicted in Figure 2(b) for an example in which the band structure has 2 and 3 valence and conduction bands, respectively. If the band structure has 100 points along the BZ, then it is required to store 600 different values for $E_{cv}$ and optical matrix elements. Now at each given value of frequency (ω), the number of $K_{zp}$ points at which $E_{cv}(K_{zp}) = \hbar\omega$ are found by counting the number of times the modified band structure, $E_{cv}(k)$, crosses the horizontal $\hbar\omega$ line [Figure 2(b)].

The absorption, α(ω), is calculated using the extinction ratio, κ(ω). The extinction ratio can be expressed in terms of real and imaginary parts of dielectric function, i.e. $\varepsilon_1(\omega)$ and $\varepsilon_2(\omega)$, respectively. Since the imaginary part of the dielectric function has already been calculated, it is possible to find the real part by applying Kramers-Krönig theorem[81]. We first show how this helps in the numerical evaluation of $\varepsilon_1(\omega)$. To show this, equation (11) is directly inserted into the Kramers-Krönig integral as follows:

$$\epsilon_1(\omega') = 1 + \frac{2}{\pi} P \left[ \int_0^\infty \frac{\varepsilon_2(\omega)\omega d\omega}{\omega^2 - \omega'^2} \right] =$$

$$1 + \frac{2}{\pi} P \left[ \int_0^\infty \frac{\pi e^2}{\epsilon_0 m_0^2 V \omega^2} \sum_{c,v} \sum_k |\langle u_c|\hat{e}.\boldsymbol{p}|u_v\rangle|^2 \delta(E_{cv}(\boldsymbol{k}) - \hbar\omega) \frac{\omega d\omega}{\omega^2 - \omega'^2} \right] =$$

$$1 + \frac{2e^2}{\epsilon_0 m_0^2 V} \sum_{c,v} \sum_k |\langle u_c|\hat{e}.\overline{\boldsymbol{P}}|u_v\rangle|^2 \int_0^\infty \delta(E_{cv}(\boldsymbol{k}) - \hbar\omega) \frac{\hbar^2 d\hbar\omega}{\hbar\omega((\hbar\omega)^2 - (\hbar\omega')^2)}. \quad (20)$$

Using the sifting property of Dirac's delta function and converting the summation over k vectors to an integration lead to:

$$\epsilon_1(\omega') = 1 + \frac{4e^2\hbar^2}{\pi\epsilon_0 m_0^2 A_{nw}} \sum_{c,v} \int_{k=0}^{k=\pi} |\langle u_c|\hat{e}.\boldsymbol{p}|u_v\rangle|^2 \frac{dk}{E_{cv}(k)((E_{cv}(k))^2 - (\hbar\omega')^2)}. \quad (21)$$

Evaluating equation (21) requires three nested summation loops. The first loop runs over the incoming light frequencies (ω'). The second loop runs over all combinations of valence to conduction band transitions (*cv*) and the third loop runs over all discrete *k* points along the BZ (i.e. $k_z$ = 0 to π). After this step, both real and imaginary parts of the dielectric function are used to find the extinction ratio (κ) and refractive index (n) of SiNW according to:

$$\kappa = \sqrt{(\sqrt{\varepsilon_1^2 + \varepsilon_2^2} - \varepsilon_1)/2} \quad \& \quad n = \sqrt{(\sqrt{\varepsilon_1^2 + \varepsilon_2^2} + \varepsilon_1)/2}. \quad (22)$$

The relation between the absorption coefficient and the extinction ratio is given as follows[81]:

$$\alpha(\omega) = \frac{4\pi\kappa(\omega)}{\lambda_0} = \frac{4\pi\kappa(\omega)}{c/f} = \frac{2\omega\kappa(\omega)}{c}. \quad (23)$$

Implementing equations (19) and (21) requires storing all combinations of valence to conduction transitions (i.e. storing the corresponding energy differences and matrix elements) which renders the fully numerical method very slow, especially for large diameter nanowires. A simpler semi-analytical formulation based on effective mass approximation (EMA) is presented in the Appendix which offers a good approximation of absorption in a computationally more efficient manner. Note that in the numerical implementation of the equations in both **B.1** and **B.2** we use the right hand side of equation (13) to find the momentum matrix elements.

**B.2 Absorption in bulk silicon:** As a validation step, we will show that tight binding method can successfully reproduce the experimental as well as DFT-based absorption spectrum of bulk silicon for all possible direct valence-to-conduction inter-band transitions.

To speed up the integration over the 3D BZ of bulk silicon, the calculation of Eigen states and absorption can be confined to an irreducible wedge as shown in Figure 3. The wedge is $1/48^{th}$ of the whole BZ, hence the calculated quantities within this wedge must be weighted correctly. The higher the number of ($k_x$, $k_y$, $k_z$) points (samples) taken from the wedge, the more precise the calculated absorption value is.

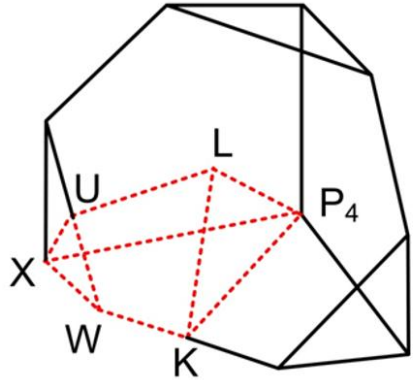

FIG. 3: The first BZ of the reciprocal lattice for bulk silicon with emphasis on the first octant (solid line) which contains the first irreducible wedge (dashed line).

To calculate the Eigen states and Eigen energies, a discretized 3D k-space is built by dividing the 3D space of $(0,2\pi/a) \times (0,2\pi/a) \times (0,2\pi/a)$ into $N_x \times N_y \times N_z$ points where *a* is the lattice constant of silicon. Then a search routine is used to find those ($k_x$, $k_y$, $k_z$) points from this space which are surrounded by the five faces of the wedge which are given by the following equations[82]:

$$k_x + k_y + k_z = \frac{3}{2}\frac{2\pi}{a}, \ k_x = k_z, k_y = k_z, k_z = \frac{2\pi}{a} \text{ and } k_z = 0 \quad (24)$$

Now $N_w$ triplets of ($k_x,k_y,k_z$) are found within the wedge which are saved for the next step of Eigen state/energy calculation. Using Bloch's theorem, we obtain the 20×20 matrix (two atoms per primitive cell), whose eigen values and functions correspond to the energy levels and wave functions[83]. The lowest 4 Eigen-states and energies belong to the valence band and the rest (16 states) compose the conduction band. Hence there are 64 possible valence to conduction band transitions at each given *k* point within the 3D irreducible wedge [dashed lines in Figure 3].

Afterwards, equation (11) is used to find the imaginary part of the dielectric function. Replacing the summation over **k** with 3D integration and using the Lorentzian broadening function to approximate the Dirac delta function, yields:

$$\epsilon_2(\omega) = \frac{\pi e^2}{\epsilon_0 m_0^2 V \omega^2} \sum_{c,v} \frac{2 \times V \times 48}{8\pi^3} \int_{\overline{K}} |\langle u_c|\hat{e}.\boldsymbol{p}|u_v\rangle|^2 \frac{\gamma/\pi}{(E_{cv}(\boldsymbol{k})-\hbar\omega)^2+\gamma^2} d^3k \qquad (25)$$

where γ is the broadening with a value of 10-20 meV chosen[70] and the volume element is $d^3k=dk_x dk_y dk_z$. Implementing equation (25) requires three nested loops corresponding to the photon energy (ω), 64 combinations of sub bands (*cv*), and $N_w$ values of ***k*** triplets within the wedge. The Kramers-Krönig code[84] is used to find the real part of the dielectric function from equation (25). The total absorption is found by averaging three values corresponding to three different polarizations (x, y, z). In parallel with the abovementioned procedure, the DFT method implemented in SIESTA® is used to calculate the dielectric function, $\epsilon_2(\omega)$, of bulk silicon. The results were post processed using available Fortran codes in SIESTA® to find $\epsilon_1(\omega)$ as well as $\alpha(\omega)$[57].

### III. RESULTS

#### A. Direct transitions in bulk silicon

The absorption spectrum of bulk silicon is calculated by TB and it is compared with experimental data[85] in Figure 4. Firstly, it is observed that the TB-based absorption spectrum starts at 3.24 eV which is consistent with the value of direct bandgap in bulk silicon. However, the DFT-based spectrum starts at a lower value due to underestimation of bandgap with the LDA functional. Secondly, the close match between the experimental data, TB and DFT results (e.g. α ≈ $10^6$ cm$^{-1}$ for $E_g$ > 3.5 eV) lends some validation for application of the tight binding method for studying the light absorption in SiNWs. The low energy tail (E < 3.24eV) of the absorption spectrum which was calculated by TB method is a numerical artifact and it has no physical meaning. This arises because the low energy skirts of many Lorentzian functions [last term of equation (25)] were being added together, and this resulted in a non-zero value. The inset of Figure 4 shows the imaginary part of dielectric function, $\epsilon_2(\omega)$, calculated by TB and DFT methods. Additionally, the static dielectric constant for bulk silicon using TB method is found to be 10.1055 in close agreement with the experimental value of 11.9 at room temperature[86] and values calculated using TB method with different number of orbitals[68].

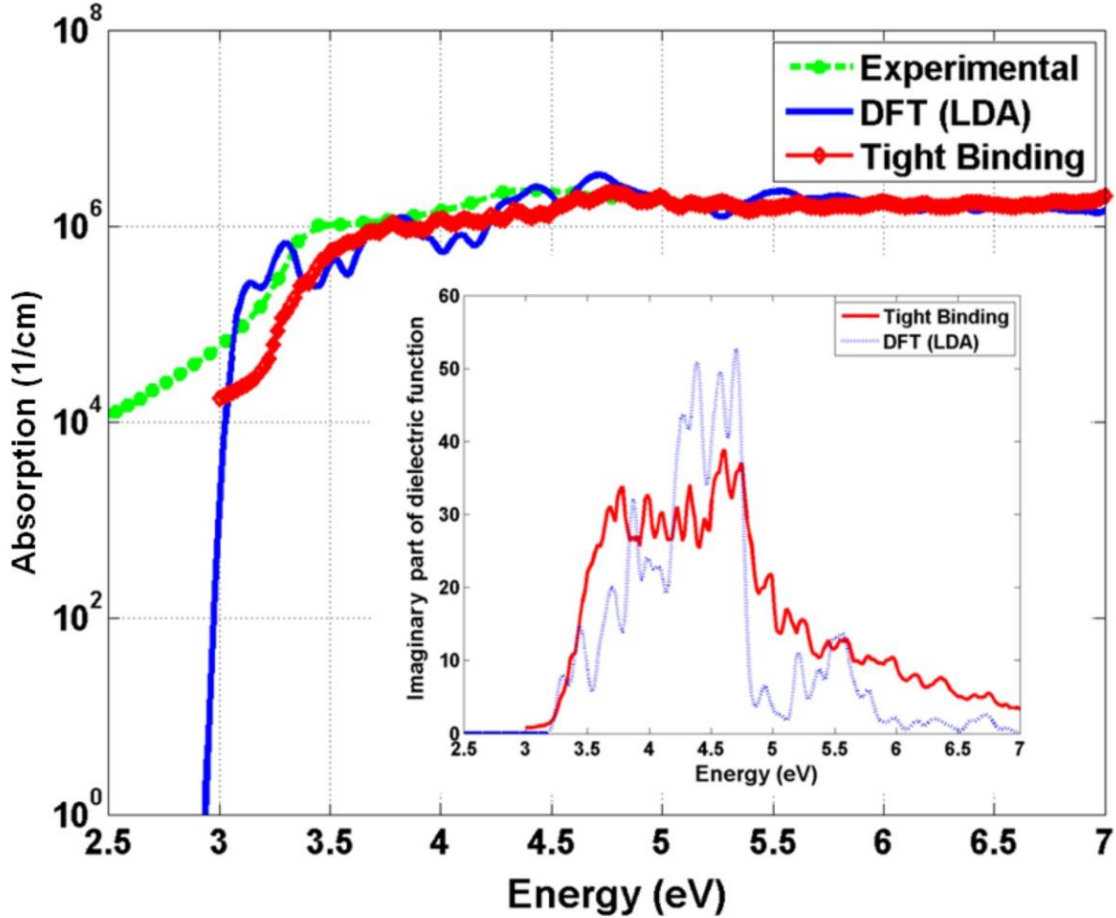

FIG. 4. Absorption spectrum of bulk silicon was calculated by DFT and TB. A comparison was made with experimental data[85]. Inset: Imaginary part of dielectric function of bulk silicon, $\varepsilon_2(\omega)$, calculated by TB and DFT methods.

B. **Absorption spectrum of [110] SiNWs**

Figure 5 compares the absorption spectra calculated using the fully numerical method (section II.B) and the semi-analytical approach based on effective mass theory (see Appendix). Using the semi-analytical method significantly enhances the computational speed at the price of losing some precision at higher energies. The calculation time is reduced to less than one tenth of the time taken by the fully numerical method. Also, it is evident that the smaller diameter nanowire (d=0.5nm) has a larger bandgap ($E_g$ = 3.6 eV) than the larger diameter one (d=1.1 nm) which has a bandgap of 2.25 eV. This is expected due to quantum confinement.

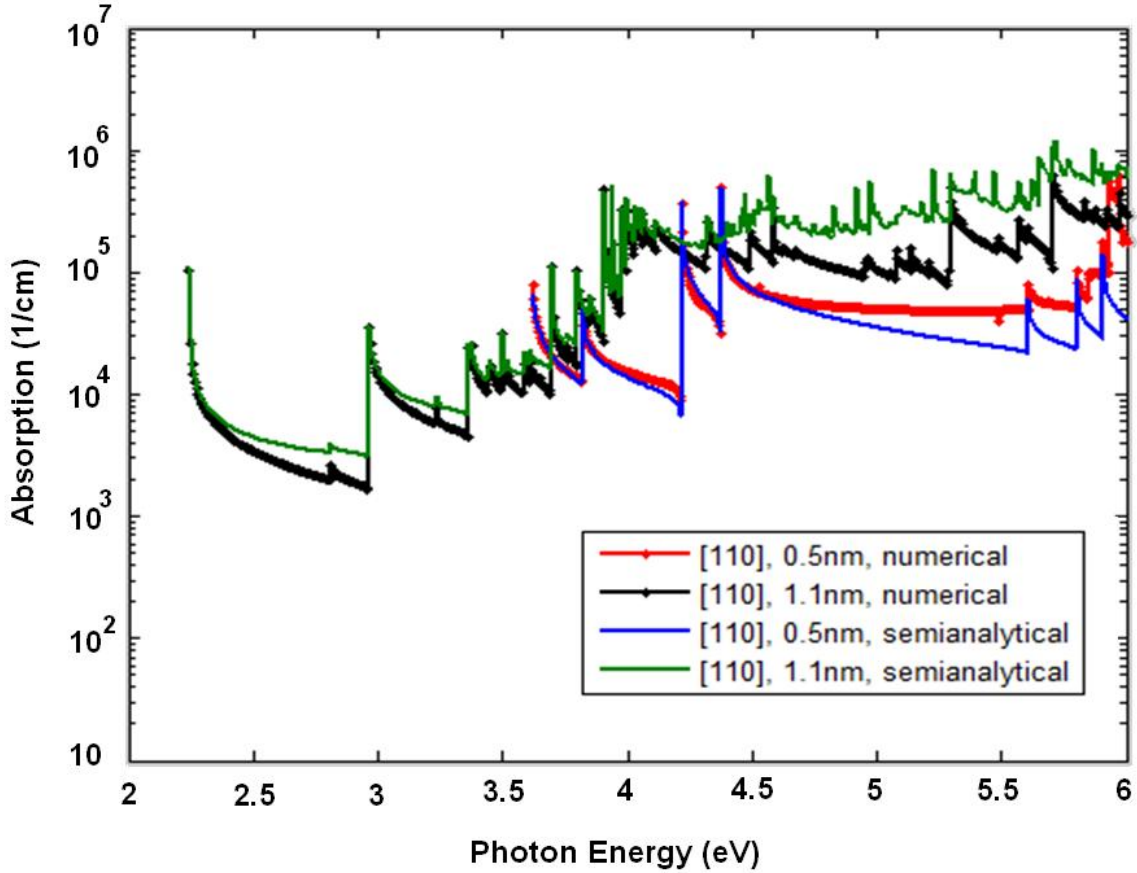

FIG. 5. Comparison of the absorption spectra for 0.5nm and 1.1nm [110] SiNWs, calculated by fully numerical and semi-analytical methods.

In terms of accuracy, it is apparent from Figure 5 that the semi-analytical approach can produce comparable results, especially for energy values below 4.5eV. This is because the $E_{cv}(k)$ sub bands fit very well to EMA-based parabola for this energy range. For higher photon energies, the sub bands tend to have a different or even negative curvature; however, some of peak positions continue to be similar.

**B.1 Effect of anisotropy and diameter:** As it was observed theoretically[45,87] and experimentally[88,89], nanowires have strong optical anisotropy. For example, in [110] silicon nanowires, the photoluminescence is stronger when polarization is parallel to the nanowire axis. The same effect is observed here in band edge absorption spectrum of [110] SiNWs. As Figure 6 shows, in a 1.7nm [110] SiNW, the absorption value for z-polarized photons is 6 orders of magnitude higher than the same quantity for x and y polarizations. In photo detectors made from SiNWs this effect could make the photocurrent highly sensitive to polarization of incident light.

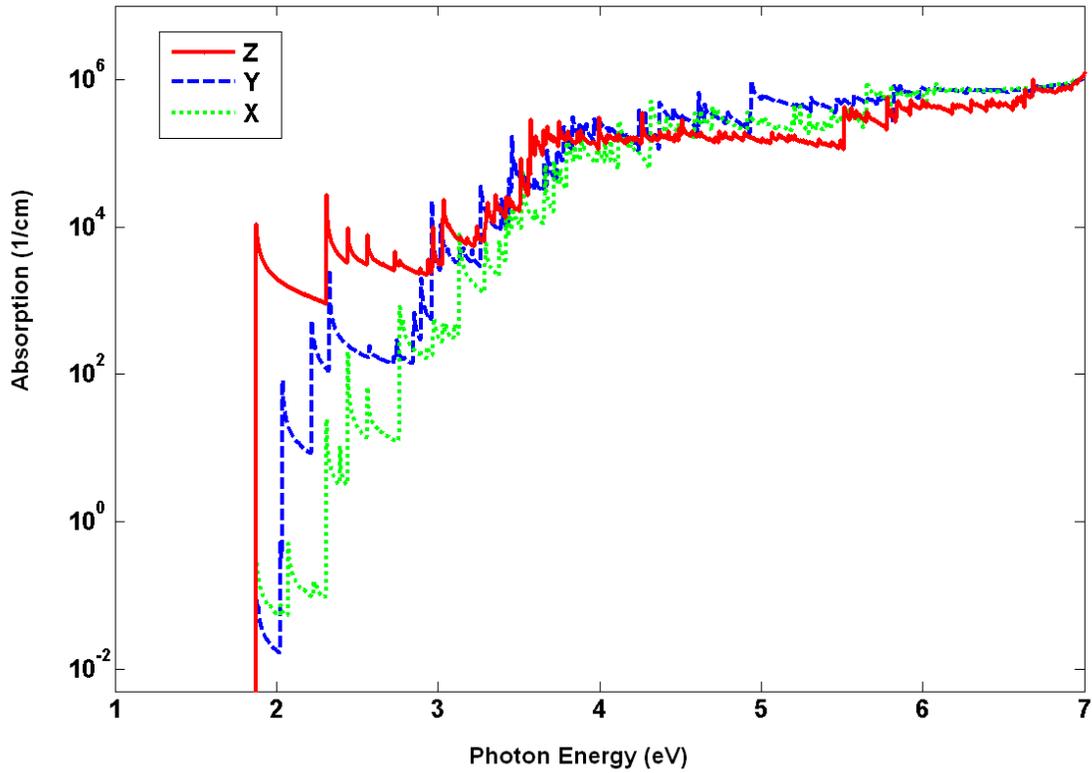

FIG. 6. Absorption spectrum of a 1.7nm [110] SiNW for the three different incident photon polarizations.

In our previous work[45], we observed that the optical dipole matrix elements for z-polarized photons had a stronger role in determining the value of spontaneous emission time in 2.3nm and 3.1nm [110] SiNWs. This means that the value of absorption for z-polarized photons is also larger for these SiNWs. We also notice that absorption is inversely proportional to diameter of the nanowire, due to decrease of optical dipole matrix element with increase in diameter (Figure 7). The change of matrix element with diameter (confinement) has been previously explained using the particle in a box model[70].

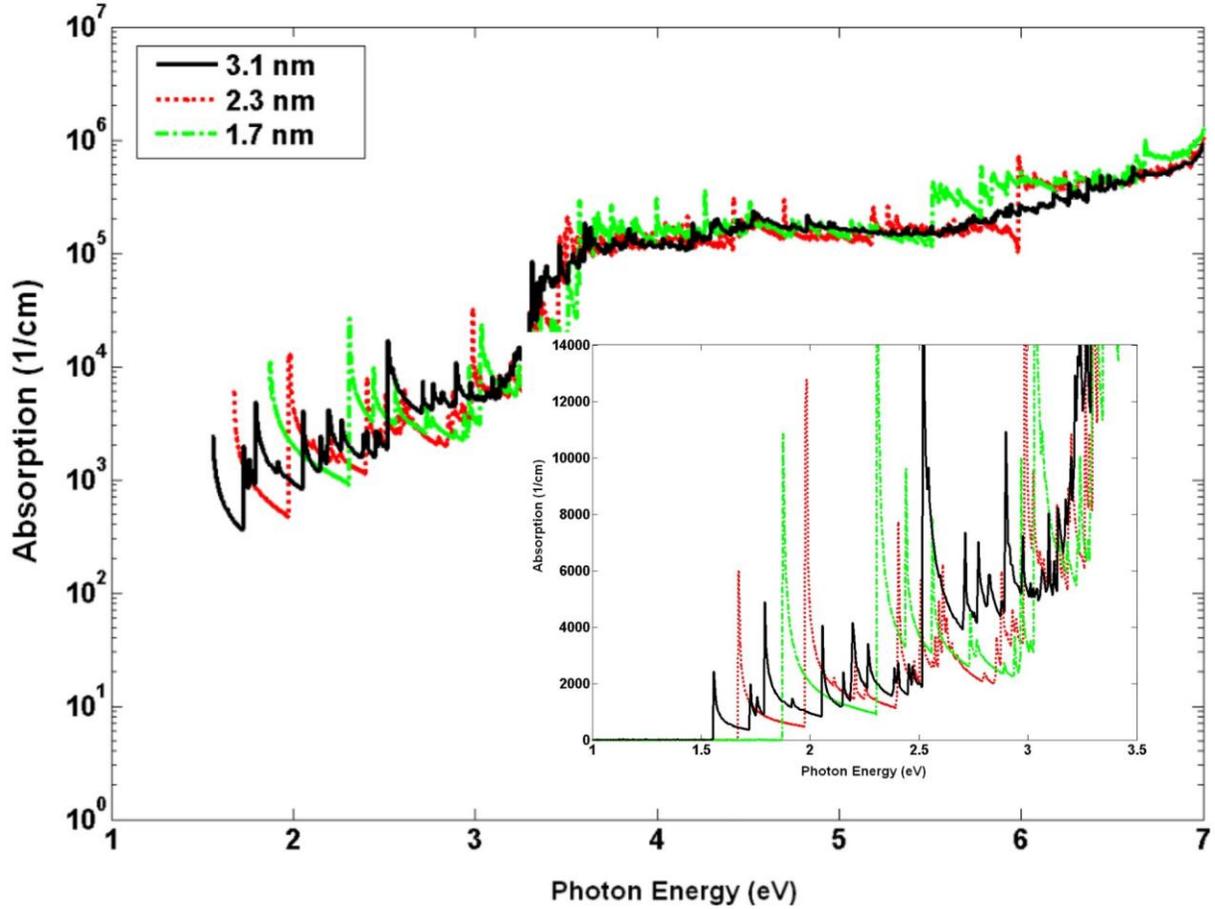

FIG. 7. Absorption spectrum of three SiNWs with diameters of 1.7nm, 2.3nm and 3.1nm. The photon polarization is along the z-axis. Inset shows the magnified section of the absorption spectra around the bandgap energy.

For the sake of clarity, the absorption spectra in Figure 7 is magnified around the band edge in order to show the exact difference between 1.7 nm, 2.3 nm and 3.1 nm [110] SiNWs. We observe in the inset that the first absorption peak belongs to a 3.1nm [110] SiNW. It has the smallest band edge absorption compared to the other nanowires considered. This result suggests that embedding arrays of silicon nanowires covering a wide range of diameters can broaden the absorption spectrum of the host material as each array is sensitive to different cut-off wavelength as a result of its bandgap.

**B.2 Effect of strain:** Effect of strain on the spontaneous emission of photons in nanowires was explained through the change of optical transition matrix element arising from the change of wave function symmetry[45]. The symmetry of wave functions was almost intact in the tensile strain regime which resulted in very close values of spontaneous emission time for this regime. On the other hand by entering into compressive strain regime the spontaneous emission time decreased by a few orders of magnitude. This is due to the valence band flip which changed the wave function symmetry.

In the context of absorption, we show that this mechanism causes a large variation of band edge absorption in the compressive strain regime. Figure 8 shows the absorption spectra of a 1.7nm [110] SiNW at -2%, 0% and +2% strain values. As it is evident, the values of absorption for 0% and +2% strained nanowires are fairly close to each other, however for the compressively strained nanowire (-2%) we observe one order of magnitude drop in the value of absorption. This is due to the valence band interchange that occurs in response to compressive strain. As shown in Figure 8, the valence sub band V2 which had lower energy than V1 in case of 0% and +2% strain, is now shifted upward as a result of its anti-bonding orbital composition in response to -2% strain[34,48,51]. Hence due to its different symmetry compared to V1, it lowers the value of optical matrix element compared with other cases in which C1 and V1 determine the band edge transition rate. The normalized momentum matrix element values for 0% and +2% strained nanowires are ~15meV at BZ center while for -2% strained nanowire it is 1.5meV. This can explain the one order of magnitude difference between absorption values at 0% and -2% strains.

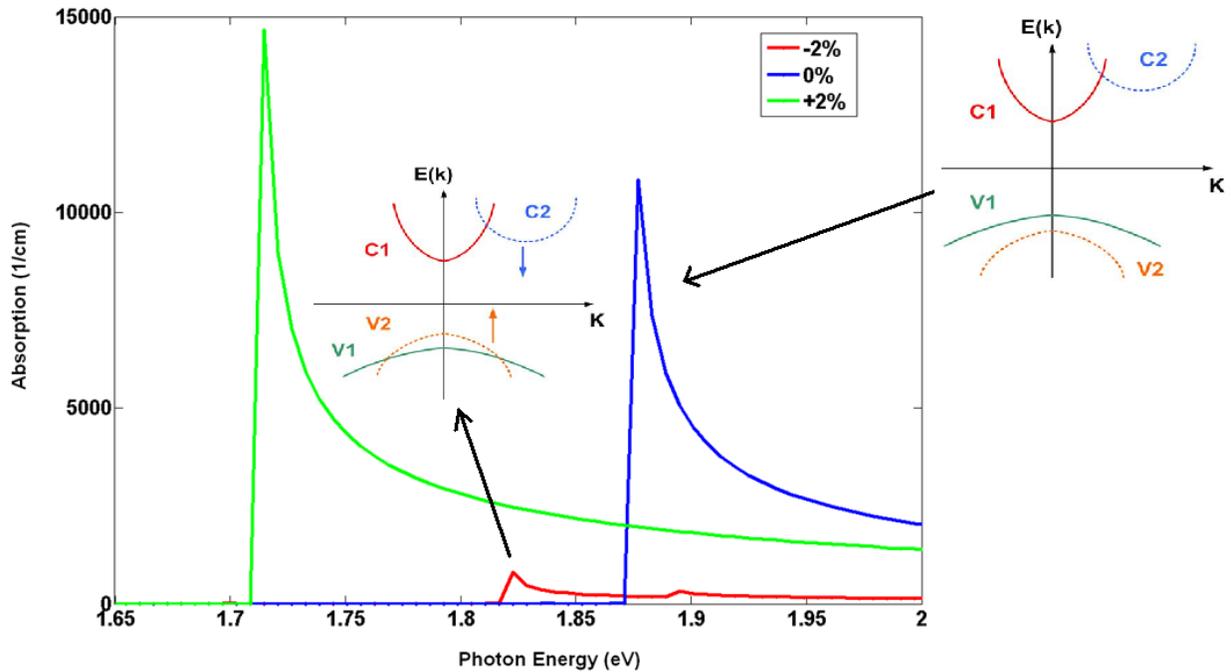

FIG. 8. Absorption spectra for -2%, 0% and +2% strained 1.7nm [110] SiNW. Incident photons are z polarized in all cases. At -2% strain, the interchange of V1 and V2 reduces the matrix element, as V2 has different symmetry than V1.

Figure 9 compares the experimental absorption spectrum of bulk silicon[85] and the calculated spectrum for a 3.1 nm [110] SiNW under +5% strain. Since the unstrained bandgap of [110] nanowires in this work are more than that of bulk silicon, a strained nanowire is chosen for comparison purpose. The nanowire under study has a direct bandgap of 1.1eV as opposed to bulk silicon which has indirect bandgap at 1.1.eV; hence, a higher absorption is expected for the nanowire provided the dipole matrix elements are non zero. It is assumed that the incident light is polarized along the z-axis. We see in Figure 9 that in the energy range of 1-2 eV, the nanowire has a greater amount of absorption than bulk silicon i.e. 7000 cm$^{-1}$ versus 100 cm$^{-1}$. Within this range, the absorption process in bulk silicon is a weak second order process mediated by phonons.

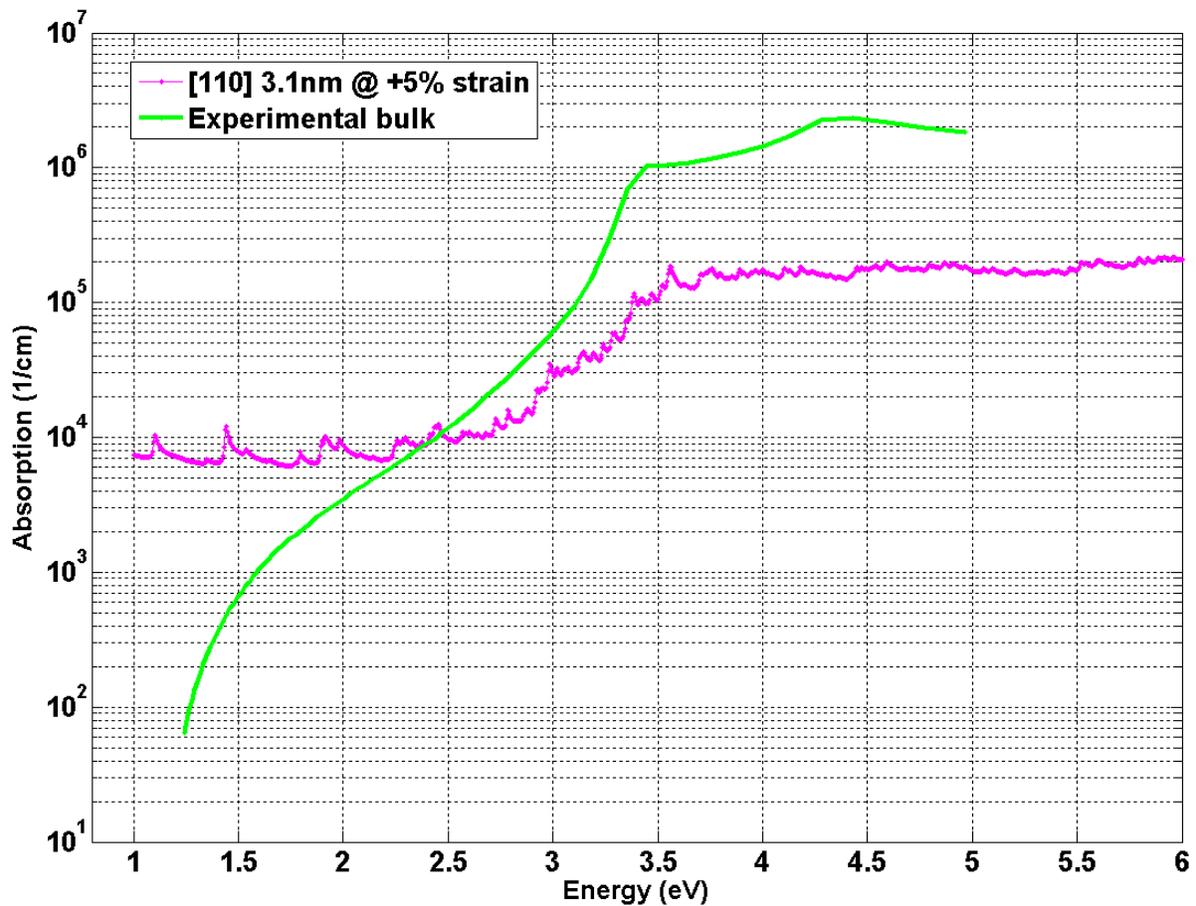

FIG. 9. Comparison of bulk silicon absorption spectrum (experimental[85] with *indirect* bandgap of 1.1 eV) with that of a 3.1 nm [110] SiNW at +5% strain (tight binding calculation with *direct* bandgap of 1.1. eV).

C. **Absorption spectrum of [100] SiNWs**

Comparing the absorption spectrum of [110] and [100] SiNWs is also instructive since it further reveals the role of wave function symmetry in determining the optical anisotropy. Figure 10 shows the absorption spectrum of a 2.2 nm [100] unstrained SiNW in a wide energy range including the magnified portion near the band edge (Inset). As it can be seen here, the band edge absorption is predominantly determined by x and y polarizations. For the z polarized case, the matrix element (and hence absorption) vanishes (Figure 10).

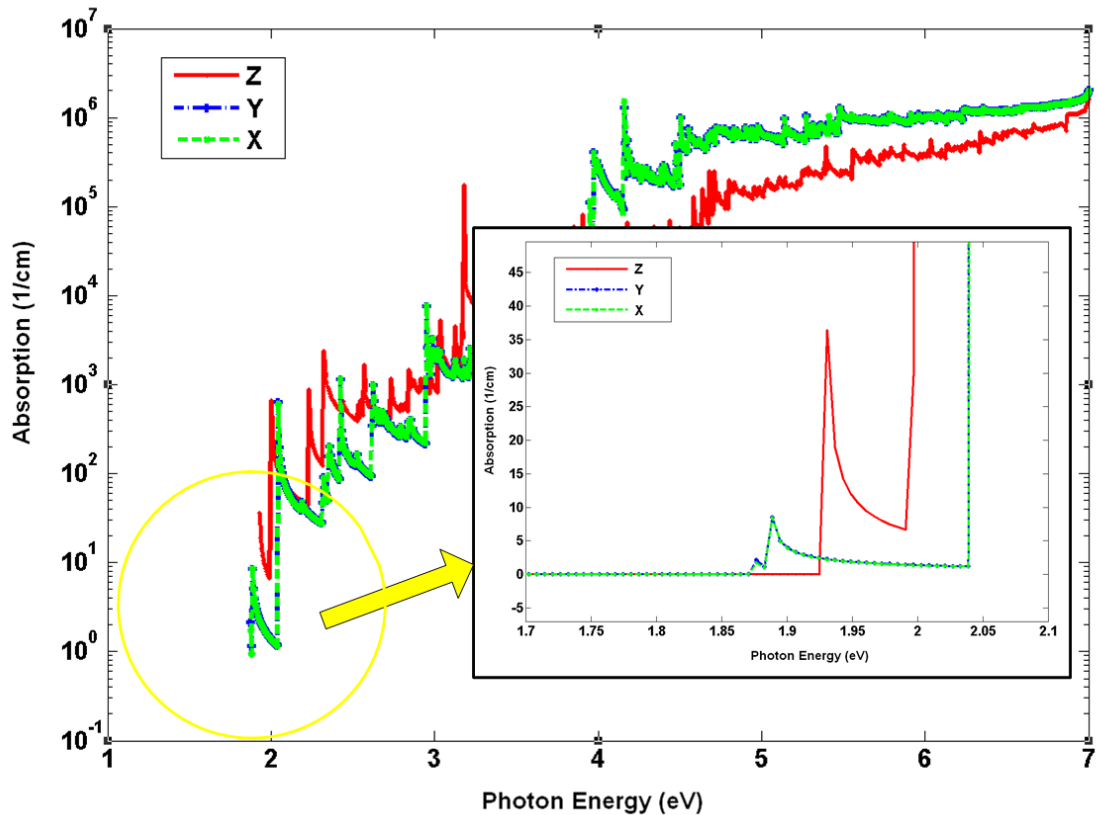

FIG. 10. Absorption spectrum of a 2.2 nm [100] SiNW. Inset shows that the band edge absorption is nonzero for x and y polarized photons while it is zero for z polarized case.

The wave functions corresponding to conduction/valence band maximum/minimum (at BZ center) are plotted in the xy cross sectional plane of the nanowire in Figure 11(a) and 11(b), respectively. The conduction and valence bands have even and odd symmetry with respect to the center of rotation, respectively. Recalling the definition of position operator matrix element, i.e. equation (1), we observe that the integrand must have an even parity in order to obtain a nonzero matrix element. Since x and y operators have odd parity, they both make the integrand to be of even symmetry. As a result, the x and y polarized matrix elements are nonzero as opposed to z-polarized matrix element. Since the cross section of the nanowire has C4 group symmetry, the z operator is of A type which results in an integrand of odd symmetry, and hence a zero value for matrix element is expected. Figure 11(c) shows the normalized momentum matrix element along the BZ of the nanowire in which x- and y- polarization components are more dominant than the z- component. Figure 11(d) shows the symmetry of the nanowire cross section, which suggests that the x- and y- components of dipole matrix elements are comparable. A similar anisotropy was observed in DFT+LDA calculations performed on a [100] SiNW with a diameter of ~1nm[87].

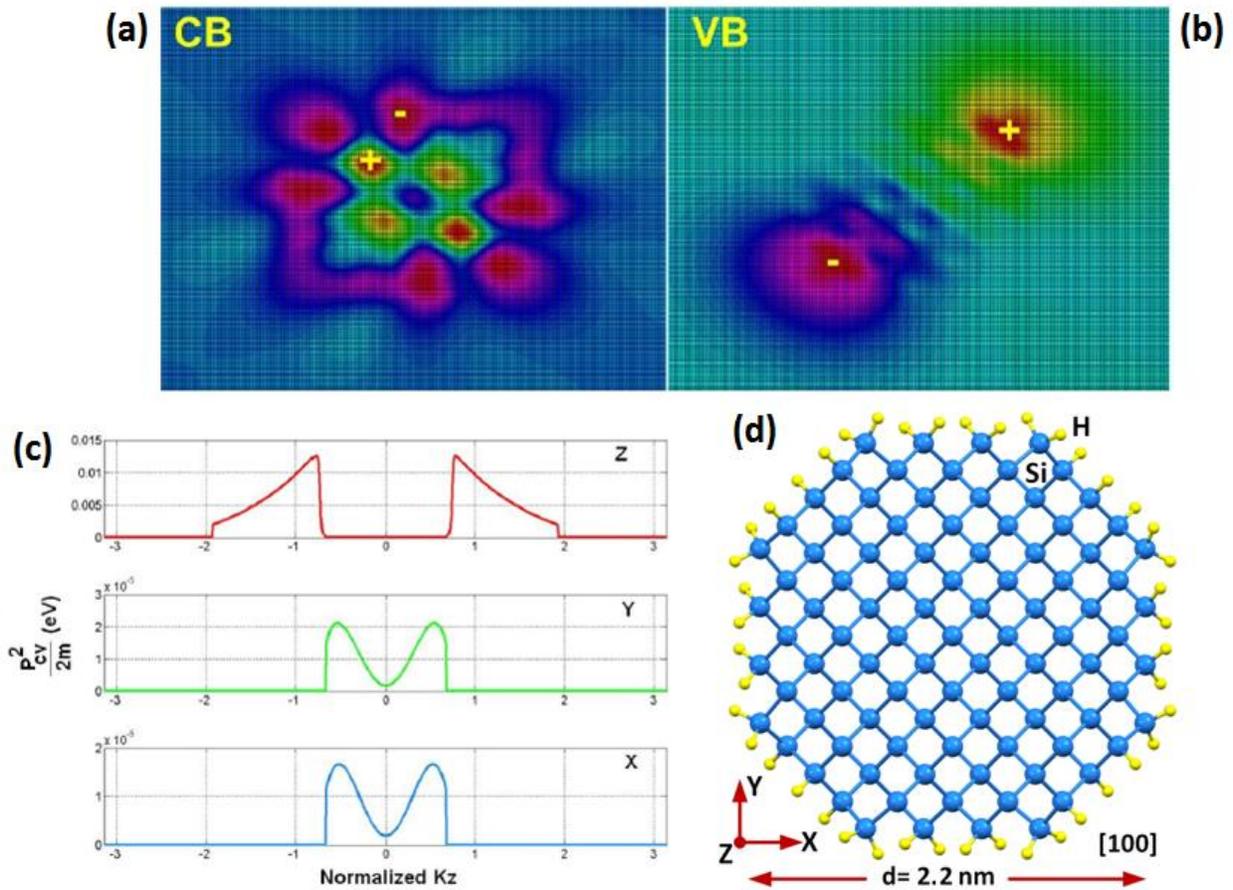

FIG. 11 Symmetry of the conduction (a) and valence (b) sub bands at BZ center reveals centro-symmetric and non centro-symmetric nature. (c) The normalized value of momentum matrix element in terms of eV along 1D BZ ($-\pi<k_z<\pi$) for x, y and z polarized photons. (d) The cross sectional view of 2.2 nm [100] SiNW. The band structure of this nanowire was shown in Figure 1(b).

## IV. CONCLUSIONS

The absorption spectra of SiNWs were calculated using semi-empirical ten orbital tight binding method. Results for the absorption spectra from this method were compared to results from DFT calculations and experimental data, for direct transitions in bulk silicon. After detailed explanations of both TB-based fully numerical and EMA-based semi-analytical methods, and demonstrating their agreement in calculating the absorption spectrum, we investigated the effect of diameter, crystallographic direction, optical anisotropy, and mechanical strain, on the absorption of SiNWs. It was observed that strain can change the band edge absorption by at least one order of magnitude due to change in the symmetry of wave functions and optical matrix element. The optical anisotropy also manifests itself in the different values of band edge absorption corresponding to different polarizations of incident photons. For the case of hexagonal cross section [110] silicon nanowires, we observed high absorption for z-polarized photons. In contrast to this, square cross section [100] nanowires showed equally strong absorption for x and y polarized photons and zero (symmetry forbidden) absorption for z polarized photons. Vital for the application of SiNWs in solar cells we found that SiNW can have 100 times larger absorption compared to bulk silicon in the same range of IR energy ($E_g$=1.1 eV). We also observed that different diameters of silicon nanowires show different band edge absorption. This study suggests potential applications of silicon nanowires in spectrum widening of solar cells or photo detectors. The equivalence of stimulated emission rate with absorption spectra facilitates the calculation of gain spectrum by assuming partially filled conduction band and partially empty valence band (population inversion). In this case, the difference of Fermi factors (last parenthesis of equation 9) must be properly updated. Based on this, tuning of emission wavelength in nanowire-based lasers using diameter, strain and crystallographic direction can be proposed.

**Acknowledgements**

Daryoush Shiri acknowledges access to the supercomputing facilities provided by Shared Hierarchical Academic Research Computing Network (SHARCNET®) in Ontario, Canada while working at University of Waterloo. The work of M. Golam Rabbani and M. P. Anantram was supported by National Science Foundation under Grant numbers 1001174 and 1231927. M. P. Anantram was also partially supported by QNRF grant (NPRP 5-968-2-403) and the University of Washington.

**APPENDIX : SEMI-ANALYTICAL CALCULATION OF ABSORPTION SPECTRUM**

Equation (21) can be simplified by making the following assumptions: *Firstly* it is assumed that the band structure is of parabolic shape. Therefore, the energy of each band can be written in effective mass formalism as:

$$E_c(k) = E_{cmin} + \frac{\hbar^2 k^2}{2m_c^*}, E_v(k) = E_{vmax} - \frac{\hbar^2 k^2}{2m_v^*} \rightarrow E_{cv}(k) = E_g + \frac{\hbar^2 k^2}{2\mu_{cv}} \quad (A1)$$

where $\mu_{cv}$ is the reduced effective mass of a conduction and valence band pair i.e. $\mu_{cv} = \frac{m_c^* m_v^*}{m_c^* + m_v^*}$.

*Secondly* it is assumed that the optical dipole matrix element value around the BZ center is almost constant. As a result, for a given combination of bands (e.g. $c_i$ and $v_j$), the optical dipole matrix element for inter band transitions at each $k$ value is equal to the matrix element value given at $k = 0$:

$$|\langle u_c|\hat{e}.\boldsymbol{p}|u_v\rangle|^2 = |\langle u_c|\hat{e}.\boldsymbol{p}|u_v\rangle|^2_{@k=0} = |\boldsymbol{p}_{cv}(k=0)|^2 \quad (A2)$$

With these assumptions, equation (19) of the main text is reduced to:

$$\epsilon_2(\omega) = \frac{2e^2}{\epsilon_0 m_0^2 A_{nw} \omega^2} \sum_{c,v} \frac{|\boldsymbol{p}_{cv}(k=0)|^2}{\left|\frac{\hbar^2 k}{2\mu_{cv}}\right|} \quad (A3)$$

The summation over $k_{zp}$ is not required here since the horizontal line of $\hbar\omega$ always intersects each $E_{cv}(k)$ at one point [See Figure 2(b)]. The crossing due to the next $E_{cv}(k)$ is already taken into account in the summation over *cv* in equation A3. This equation can be simplified by replacing *k* with the following value:

$$k = \frac{\sqrt{2\mu_{cv}}}{\hbar}\sqrt{E_{cv}(k) - E_{g,cv}} = \frac{\sqrt{2\mu_{cv}}}{\hbar}\sqrt{\hbar\omega - E_{g,cv}}, \quad (A4)$$

where $E_{g,cv}$ represents the minimum of $E_{cv}(k)$ [i.e. $E_{cv}(k=0)$] for the corresponding *cv* index. For example in Figure 2(b), $E_{g,cv}$ spans six values; hence equation A4 is further reduced to the following equation in which the summation adds the six terms sequentially.

$$\epsilon_2(\omega) = \frac{2e^2}{\epsilon_0 m_0^2 A_{nw} \omega^2} \sum_{c,v} \frac{|\boldsymbol{p}_{cv}(k=0)|^2 \sqrt{2\mu_{cv}}}{\hbar\sqrt{\hbar\omega - E_{g,cv}}} = \frac{2\sqrt{2}e^2}{\epsilon_0 m_0^2 A_{nw} \omega^2 \hbar} \sum_{c,v} \frac{|\boldsymbol{p}_{cv}(k=0)|^2 \sqrt{\mu_{cv}}}{\sqrt{\hbar\omega - E_{g,cv}}} \quad (A5)$$

For each photon energy $\hbar\omega$, the summation runs only over all combinations of conduction and valence band (*cv*) with $E_c - E_v = \hbar\omega$ and optical matrix elements at $k = 0$. Therefore, this method is advantageous over the fully numerical method [equation (19)] if it can show that both methods give similar results. To obviate the singularities at $\hbar\omega = E_{g,cv}$, we can use the following approximation by introducing a broadening factor[70] of η:

$$\frac{1}{\sqrt{x}} \cong \frac{1}{\sqrt{x+i\eta}} = \frac{\sqrt{x-i\eta}}{\sqrt{x^2+\eta^2}} \cong real\left(\frac{\sqrt{x-i\eta}}{\sqrt{x^2+\eta^2}}\right) \quad (A6)$$

The real part of the dielectric function is found by inserting equation A5 into the Kramers-Krönig integral which yields:

$$\epsilon_1(\omega) = 1 + \frac{4\sqrt{2}e^2}{\pi\epsilon_0 m_0^2 A_{nw}\omega^2 \hbar}\sum_{cv}|\boldsymbol{P}_{cv}(k=0)|^2\sqrt{\mu_{cv}}\, P\left[\int_0^\infty \frac{d\omega'}{\omega'(\omega'^2-\omega^2)\sqrt{\omega'-(E_{g,cv})/\hbar}}\right] \quad (A7)$$

The right hand side of equation A7 can be further simplified using Wolfram Mathematica® online integrator[79] or the code available in Ref. 84, and using the Kramers-Krönig integral. Calculating extinction ratio and absorption are then completed using equations (22) and (23).